\documentstyle[aps,pra,manuscript]{revtex}   
\begin{document}
\draft
\title{Modified Bloch equations in presence of a nonstationary bath}
\author{Jyotipratim Ray Chaudhuri, Suman Kumar Banik,
Bimalendu Deb{\footnote   
{present address : Department of Chemical Physics, Weizmann
Institute of Science, Rehovot, Israel.}} and 
Deb Shankar Ray{\footnote {e-mail: pcdsr@mahendra.iacs.res.in}} }
\address{Indian Association for the Cultivation of Science,
Jadavpur, Calcutta 700032, INDIA.}
\maketitle
\begin{abstract}
Based on the system-reservoir description we propose a simple solvable
microscopic model for a nonequilibrium bath.
This captures the essential features of a nonstationary
quantum Markov process. We establish an appropriate generalization of the 
fluctuation-dissipation relation pertaining to this process and explore the 
essential modifications of the Bloch equations to reveal the nonexponential
decay of the Bloch vector components and transient spectral broadening in 
resonance fluorescence. We discuss a simple experimental scheme to verify
the theoretical results.
\end{abstract}

\vspace{2.0cm}



\newpage

\section{Introduction}

\vspace{0.5cm}

The dynamics of most of the quantum optical phenomena are based on two
fundamental processes; the coherent interaction between the
system (atom/molecule) and the field mode (classical/quantum) 
and the incoherent dissipation
of the system. The latter is traditionally modeled in terms of the wellknown
system-reservoir theory within the appropriate finite temperature quantum
statistical scheme \cite{louisell,leggett,dsr,lu,gardiner}. 
Besides thermal reservoirs, the non-thermal reservoirs 
\cite{lu,gardiner} have also been 
found to be important in connection with the development of correlated emission
lasers \cite{lu} and squeezed light fields \cite{gardiner}. The essential underlying assumption about 
the bath, be it thermal or nonthermal, is that it is considered to be in a 
state of equilibrium throughout the process. Very recently a solvable 
microscopic model for a nonequilibrium bath  
has been proposed \cite{jcp} to explore classically, the influence of an initial 
nonequilibrium excitation in a complex system on the relaxation of a specific
quantity of interest. In the present paper we extend this treatment to a quantum
optical context. Since the initial excitation creates a nonequilibrium
energy density fluctuation distribution which imparts 
nonstationarity of the bath, it is expected that optical
Bloch equations which take into account of both the coherent interaction
and the relaxation processes within a simplified description of a two-level
scheme, are likely to be modified by the nonstationarity of the bath \cite{jcp,land}.
Based on a quantum version of the model
we study this essential modification of the optical Bloch equations and 
explore some of the relevant consequences. 

We thus consider a two-level system in contact with a bath
which is not in a thermal Boltzmann distribution. 
This nonequilibrium bath is effectively realized  in terms of a semi-infinite
dimensional broad-band reservoir which is subsequently kept in contact with
a standard thermal bath which allows the nonthermal bath to relax with a characteristic
time scale. The important separation of the time scales of fluctuations of
the nonequilibrium and the thermal bath is that \cite{jcp}
the former remains effectively stationary on the fast correlation of the
thermal noise. The model captures the essential features of a
{\it nonstationary quantum Markov process}. The physical situation that
has been addressed is the following. At $t=0$ the excitation is switched on
and the intermediate bath is thrown into a nonstationary state. We then 
follow the coherent
dynamics of a classical laser-driven near-resonant two-level system 
interrupted by incoherent emissive processes due to nonequilibrium 
intermediate modes 
after $t > 0$ to observe the influence of relaxation of these modes on
the transient characteristics of the system. We show that the decay of the 
Bloch vector components is nonexponential in character so long as the
nonstationarity persists. In addition, the nonstationarity of the bath results 
in time-dependence of the diffusion coefficient which manifests itself in the
transient resonance fluorescence spectra of the two-level system. 
The underlying physical
mechanism of the transient characteristics can be understood with the help
of a generalized nonequilibrium fluctuation-dissipation relation pertaining
to this nonstationary quantum Markov process. In the long time limit one,
however, recovers the standard Bloch equations and the spectral features.

The outline of the paper is as follows : In Sec.II we discuss the model for
nonequilibrium bath and the generalization of the fluctuation-dissipation 
relation corresponding to the nonstationary process. The application of the 
model to explore the modification of the optical Bloch equations and the 
transient spectral characteristics of resonance fluorescence have been 
carried out in Sec.III. In Sec.IV we propose a simple experimental scheme to
verify the theoretical results on nonexponential decay and transient
broadening effect. The paper is concluded with a summary of the main results.

\vspace{0.5cm}

\section{
Relaxation of a two-level atom in presence of a nonstationary bath}

\vspace{0.5cm}

To start with we consider a model two-level atom (the system) coupled to a set 
of relaxing modes considered as a semi-infinite dimensional system which
effectively constitutes a nonequilibrium reservoir. This, in turn, is in
contact with a thermally equilibrated reservoir. Both the reservoirs are
composed of two sets of harmonic oscillators characterized by the frequency
sets $\{\omega_j\}$ and $\{\Omega_j\}$ for the equilibrium and nonequilibrium
bath, respectively. The total Hamiltonian is given by
\begin{eqnarray}
H_0 & = & \frac{1}{2} \hbar \omega_0 \sigma_z + \hbar \sum_j \omega_j b_j^\dagger
b_j + \hbar \sum_\mu \Omega_\mu c_\mu^\dagger c_\mu \nonumber\\ 
& + & \hbar \sum_\mu g_\mu ( \sigma_+ c_\mu + \sigma_- c_\mu^\dagger ) 
+ \hbar \sum_j \sum_\mu \alpha_{j\mu} ( b_j^\dagger c_\mu + 
b_j c_\mu^\dagger ) \; \; .
\end{eqnarray}

\noindent
The Hamiltonian is essentially a simpler quantum version of the model used 
in \cite{jcp} with
two-level atom as the system. The first term on the right-hand side describes 
the system mode with 
characteristic frequency $\omega_0$. The second and the third terms represent 
the thermal and the nonequilibrium linear modes. The next two terms represent 
the coupling of the nonequilibrium bath with the system mode and the thermal
bath where the coupling constants are $g_{\mu}$ and $\alpha_{j \mu}$, 
respectively. In writing down the Hamiltonian we have made use of the
rotating wave approximation.

The Heisenberg equations of motion for the system and the reservoir operators
are given by
\begin{eqnarray}
{\dot{\sigma}}_+ (t) & = & i \omega_0 \sigma_+ - i\sum_\mu g_\mu c_\mu^\dagger
\sigma_z  \; \; , \\
{\dot{\sigma}}_- (t) & = & -i \omega_0 \sigma_- + i\sum_\mu g_\mu c_\mu 
\sigma_z  \; \; , \\
{\dot{\sigma}}_z (t) & = & -2i \sum_\mu g_\mu c_\mu \sigma_+ + 2i \sum_\mu
g_\mu c_\mu^\dagger \sigma_- \; \; , \\
{\dot{b}}_j (t) & = & -i\omega_j b_j - i \sum_\mu \alpha_{j\mu } c_\mu \; \; , \\
{\dot{c}}_\mu (t) & = & -i\Omega_\mu c_\mu - i g_\mu \sigma_- - i\sum_j
\alpha_{j\mu } b_j \; \; .
\end{eqnarray}

\noindent
Making use of the formal integral of Eq.(5) for $b_{j}(t)$ in Eq.(6) we
obtain
\begin{eqnarray}
{\dot{c}}_\mu (t) = & - & i \Omega_\mu c_\mu - i g_\mu \sigma_- - i\sum_j
\alpha_{j\mu } e^{-i \omega_j (t-t_0)} b_j (t_0) \nonumber\\
& - & \sum_j \sum_\nu \alpha_{j\mu } \alpha_{j\nu } \int _{t_0}^t dt' c_\nu
(t') e^{-i\omega_j (t-t')} \; \; .
\end{eqnarray}

\noindent
Taking into consideration that the interference time of
$\sum_{j}\alpha_{j \mu}\alpha_{j \nu} e^{-i \omega_{j}(t-t')}$ is much
smaller than the time
over which the significant phase and amplitude modulation of the linear modes
$c_{\mu}(t)$ takes place, the last term in Eq.(7) can be identified as a
relaxation term in the usual way \cite{louisell} with the damping constant
\begin{equation}
\gamma_{\mu\nu}^c = \pi \; \alpha_{\nu\mu} (\Omega_\nu) \; \alpha_{\nu\nu}
(\Omega_\nu) \; D (\Omega_\nu) \; \; ,
\end{equation}

\noindent
where $D(\Omega_\nu)$ represents the density of states of the equilibrium
modes evaluated at $\Omega_\nu$. Thus one can write down the Langevin equation
for the relaxing mode $c_\mu$ as follows;
\begin{equation}
{\dot{c}}_\mu (t) = -i\Omega_\mu c_\mu (t) - i g_\mu \sigma_- (t) - 
\sum_\nu \gamma_{\mu\nu}^c c_\nu (t) + f_\mu (t) \; \; .
\end{equation}

\noindent
Here the last term $f_{\mu}(t)$ represents the usual noise operator arising 
out of the coupling of the relaxing modes with the thermal bath modes as 
given by
\begin{equation}
f_\mu (t) = -i \sum_j \alpha _{j\mu} e^{-i\omega_j (t-t_0 )} b_j (t_0) \; \; ,
\end{equation}

\noindent
where the reservoir average of $f_{\mu}(t)$ is zero, i.e.,
\begin{equation}
\langle f_\mu (t) \rangle_B = 0 \; \; .
\end{equation}

Here by average $\langle O(t)\rangle_B$ of an operator $O(t)$ we mean
\begin{eqnarray*}
\langle O(t)\rangle_B = Tr \{ O(t) \rho_B\} \; \; ,{\rm where}\; \;
\rho_B = \Pi_j \; \exp\{(-\hbar \omega_j b_j^\dagger b_j)/KT\}
/[1-\exp(-\frac{\hbar\omega_j}{KT})] \; \; .
\end{eqnarray*}
\noindent
$\rho_B$ is the thermal operator for initial density matrix for the
thermal bath ( initial density for the thermal bath $\{ b_j\}$,
nonequilibrium bath $\{c_\mu\}$  and the system are assumed to be
factorizable ).

We now make the following approximations. The cross-terms which involve
rapidly evolving imaginary exponentials in the summation among the bath
modes in Eq.(9) are neglected with respect to the diagonals, slowly evolving
terms. This secular approximation is the usual one made in the context of 
master equations for baths \cite{louisell}. It is wellknown  
\cite{louisell} that this approximation
is valid in the limit of weak coupling ( 
$|\alpha_{\mu\nu}| \ll |\Omega_{\nu}|$ ) and of a `flat' bath spectrum
for which $|\alpha_{\mu\nu}| \approx |\alpha_{\mu'\nu'}|$. 

Taking into consideration of the above approximations the
Langevin equation for the relaxing modes, Eq.(9), can be written in the 
following form,
\begin{equation}
{\dot{c}}_\mu (t) = -i\Omega_\mu c_\mu (t) - ig_\mu \sigma_- (t) -
\gamma_{\mu\mu}^c c_\mu (t) + f_\mu (t) \; \; .
\end{equation}

To explore the influence of an initial excitation of the intermediate 
reservoir and its relaxation, we now consider the evolution of these 
linear modes $c_\mu$ in terms of Eq.(12), which allows a formal solution
of the following form \cite{jcp}
\begin{equation}
c_\mu (t)  =   c_\mu^s (t) + c_\mu (t_0) \; 
e^{ (-i\Omega_\mu - \gamma_{\mu\mu}^c ) (t-t_0)} 
 -  i g_\mu \int_{t_0}^t dt' \; 
e^{ (-i\Omega_\mu - \gamma_{\mu\mu}^c ) (t-t')} \; \sigma_- (t') \; \; .
\end{equation}

\noindent
The first term on the right hand side in the absence of the coupling of the 
system mode represents the ( long time ) stationary stochastic solution of the 
form
\begin{equation}
c_\mu^s (t) = c_\mu^s \; e^{-i[ \Omega_\mu (t-t_0) + \phi_\mu^s]} \; \; ,
\end{equation}

\noindent
where the amplitude $c_\mu^s$ (operator) and the phases $\phi_\mu^s$
(c-number) are assumed to be randomly distributed \cite{jcp}. The random
distribution of phases and amplitudes in the stationary regime makes Eq.(13)
an instantaneous solution. The second term on the right hand side in Eq.(13)
carries the information of relaxation of the $c_\mu$ modes due to their
coupling to the thermal bath and is in the form of a typical "memory-type
term" [see the discussion later].
The latter is not to be confused with the usual memory term ( or kernel )
commonly arising out of the frequency dependence of friction. The third term 
on the other hand represents the effect of coupling of the system mode to
the nonequilibrium reservoir.

We now substitute this solution (13) in Eq.(2) to obtain the equation of 
motion for the system operator in the usual way as
\begin{equation}
{\dot{\sigma}}_+ (t) = i \omega_0 \sigma_+ (t) - \Gamma \sigma_+ (t) +
Z^\dagger \sigma_z (t)  \; \; ,
\end{equation}

\noindent
with
\begin{equation}
\Gamma = \pi \; g^2 (\omega_0) \; \rho (\omega_0) \; \; ,
\end{equation}

\noindent
where the $\rho (\Omega)$ represents the density of relaxing intermediate
oscillator modes. We assume further the weak dependence of 
$\gamma^c_{\mu\mu}$ on the modes to perform the integration over $\Omega$.

\noindent
$\Gamma$ can be identified as a dissipation constant of the system mode due 
to the fluctuation of these modes. Also note that 
\begin{eqnarray*}
Z^\dagger (t) = -i \sum_\mu g_\mu \; \left [ \; 
c_\mu^s (t) + c_\mu^\dagger (t_0)
\; e^{(i\Omega_\mu- \gamma_{\mu\mu}^c ) (t-t_0)} \; \right ]
\end{eqnarray*}

\noindent
is the noise operator for the nonequlibrium bath modes with 
$\langle Z^\dagger (t) \rangle_{NR} =0$. Here by $\langle O(t)\rangle_{NR}$
we mean
\begin{eqnarray*}
\langle O(t)\rangle_{NR} = Tr \{ O(t)\rho_c\} \; , \; \; {\rm where} \; \; \;
\rho_c = \Pi_\mu \exp\{(-\hbar \Omega_\mu c_\mu^\dagger c_\mu)/KT\}
/[1-\exp(-\frac{\hbar\Omega_\mu}{KT})] \; \; .
\end{eqnarray*}
\noindent
Here $\rho_c$ is the initial thermal density operator for the nonequilibrium
bath.

We proceed similarly to obtain the other equations of motion for system 
operators, $\sigma_-$ and $\sigma_+$ as,
\begin{eqnarray}
{\dot{\sigma}}_- (t) & = & -i \omega_0 \sigma_- (t) - \Gamma \sigma_- (t)
+ Z (t) \sigma_z \; \; ,\\
{\dot{\sigma}}_z (t) & = & -2\Gamma (1+\sigma_z) + 2 Z(t) \sigma_+ (t) +
2 Z^\dagger(t) \sigma_- (t) \; \; .
\end{eqnarray}

Introducing the slowly varying operators as
\begin{equation}
\left. \begin{array}{ccc}
{\tilde{S}}_+ (t) & = & \sigma_+ (t) \; e^{-i\omega_0 (t-t_0)} \\
{\tilde{S}}_- (t) & = & \sigma_- (t) \; e^{i\omega_0 (t-t_0)} \\
{\tilde{S}}_z (t) & = & \frac{1}{2} \sigma_z
\end{array} \right \} \; \; ,
\end{equation}

\noindent
we obtain the following Langevin equations,
\begin{equation}
\left. \begin{array}{ccc}
{\dot{\tilde{S}}}_+ (t) & = & -\Gamma {\tilde S}_+ (t) + 2{\tilde {\xi}}
^\dagger (t) {\tilde{S}}_z (t) \\
{\dot{\tilde{S}}}_- (t) & = & -\Gamma {\tilde{S}}_- (t) + 2{\tilde{\xi}} 
(t) {\tilde{S}}_z (t) \\
{\dot{\tilde{S}}}_z (t) & = & -2\Gamma [\frac{1}{2} + {\tilde{S}}_z (t) ] + 
{\tilde{\xi}}^\dagger (t) {\tilde{S}}_- (t)+ {\tilde{\xi}} (t) 
{\tilde{S}}_+ (t) \end{array} \right \} \; \; ,
\end{equation}

\noindent
where
\begin{equation}
{\tilde{\xi}}^\dagger (t) = - i \sum_\mu g_\mu \left [ c_\mu^{s \dagger} (t)
e^{-i\omega_0 (t-t_0)} + c_\mu^\dagger (t_0) e^{i(\Omega_\mu - \omega_0 )
(t-t_0)} \; e^{-\gamma_{\mu\mu}^c (t-t_0)} \right ] \; .
\end{equation}

The nonequilibrium generalization of the fluctuation-dissipation relation is 
now immediately apparent. Using Eq.(21) we have
\begin{eqnarray}
\langle {\tilde{\xi}}^\dagger (t) {\tilde{\xi}} (t')  \rangle_{NR} & = & 
\sum_\mu g_\mu^2
\left [ \langle c_\mu^{s\dagger} c_\mu^s \rangle_{NR} \;
e^{i(\Omega_\mu - \omega_0 ) (t-t')} \right. \nonumber \\
& & \left .+ \langle c_\mu^\dagger (t_0) c_\mu (t_0) \rangle_{NR} \;
e^{i(\Omega_\mu - \omega_0) (t-t')} \; e^{2\gamma_{\mu\mu}^c t_0} \;
e^{-\gamma_{\mu\mu}^c (t+t') } \;  \right ] \; \; .
\end{eqnarray}

\noindent
We denote the average photon number of the nonequilibrium bath by
\begin{equation}
{\bar{n}} (\Omega_\mu , t_0) = \langle c_\mu^\dagger (t_0) c_\mu (t_0) 
\rangle_{NR} \; \; ,
\end{equation}

\noindent
where $t_0$ signifies the dependence of average photon number of the
nonequilibrium bath on its initial state of preparation. Also the steady
state average photon number is given by
\begin{eqnarray*}
{\bar{n}} (\Omega_\mu) = \langle c_\mu^{s\dagger} c_\mu^s \rangle_{NR} \; \; .
\end{eqnarray*}

\noindent
After replacing the summation by integration and $\gamma_{\mu \mu}^c$ by an
average $\gamma$ in Eq.(22) we obtain in the usual way
\begin{equation}
\langle {\tilde{\xi}}^\dagger (t) {\tilde{\xi}} (t') \rangle_{NR} = \left [ 
\Gamma \;
{\bar{n}} (\omega_0) + e^{-2\gamma (t-t_0) } \; \Gamma {\bar{n}} (\omega_0, t_0
) \right ] \delta (t-t') \; \; .
\end{equation}

Eq.(24) and $\langle {\tilde{\xi}}(t) \rangle = 0$ summarizes the essential
properties of the stochastic processes due to intermediate oscillator bath
modes $\{c_\mu\}$. It is important to emphasize that the exponential term
in Eq.(24) [$\exp \{-2\gamma(t-t_0)\}$] does not contain time-difference
of the two different instants $t$ and $t'$ over which the stochastic
process is correlated. Thus this exponential term is not to be confused with
$\exp \{-2\gamma(t-t')\}$ which normally appears as a typical memory term
in correlation function as
\begin{eqnarray*}
\langle \xi(t) \xi(t') \rangle = {\rm Constant} \times
\exp [-2\gamma(t-t')] \; \;
\end{eqnarray*}

\noindent
in a non-Markovian stochastic process. The time-difference of the two instants
$t$ and $t'$ in the present study appear as an argument of a delta function
[$\delta(t-t')$] in Eq.(24) due to the use of standard broad-band
reservoirs. $t$ in the exponential function in Eq.(24) is rather a slow time
variable, which makes the stochastic process due to $\{c_\mu\}$-modes
nonstationary. In other words the correlation function in Eq.(24) is not
invariant under time translation. We are thus concerned here with a
{\it nonstationary quantum Markov process}. This consideration is essential for the
application of Onsager's regression hypothesis for calculation of spectra
with two-time correlation function in the present case as shown in the next
section.

\noindent
Rewriting $\Gamma {\bar{n}}(\omega_0, t_o)$ in Eq.(24) in terms of a 
deviation from its steady state value $\Gamma {\bar{n}}(\omega_0)$ as
\begin{eqnarray*}
\Gamma {\bar{n}} (\omega_0, t_0) = D(t_0) - \Gamma {\bar{n}}(\omega_0) \; \; ,
\end{eqnarray*}

\noindent
we identify a time-dependent diffusion coefficient $D(t)$ in the last 
equation (24) as
\begin{equation}
D(t) = \Gamma {\bar{n}} (\omega_0) + \left [ D(t_0) - \Gamma {\bar{n}}
(\omega_0) \right ] e^{-2\gamma(t-t_0) } \; \; .
\end{equation}

\noindent
We thus obtain 
\begin{equation}
\langle {\tilde{\xi}}^\dagger (t) {\tilde{\xi}} (t')
\rangle_{NR}  =  
\Gamma {\bar{n}}(\omega_0) \left [ 1+ r e^{-2\gamma(t-t_0)} \right ] \; 
\delta (t-t') \; \; ,
\end{equation}

\noindent
where we denote 
\begin{equation}
\left. \begin{array}{lll}
\Gamma{\bar{n}} (\omega_0) & = & D(\infty) \\
r & = & \frac{D(t_0)}{D(\infty)} - 1  \\
& = & \frac{ {\overline{n}}(\omega_0,t_0) }{ {\overline{n}}(\omega_0) }
\end{array} \right \}\; \; .
\end{equation}

Eq.(26) is the desired nonequilibrium quantum generalization of the 
fluctuation-dissipation relationship. The classical version of the above 
equation is given in \cite{jcp}. This relates instantaneous fluctuations of the
nonequilibrium bath (which itself is undergoing relaxation at a rate $\gamma$
due to its coupling with the thermal bath ) to the dissipation of the energy
of the system mode through $\Gamma$. The nonequilibrium nature of the bath
is implicit in the initial preparation which creates an initial diffusion
coefficient $D(t_0)$ and also in the exponentially decaying term. 

To check the consistency of the treatment and to allow ourselves a fair 
comparison with the classical treatment we now make the following comments.

\noindent
(i) In the steady state limit one recovers the usual fluctuation-dissipation 
relation for a thermal bath at equilibrium.

\noindent
(ii) Eq.(26) can also be expressed in terms of energy density fluctuations
of the nonequilibrium modes. The energy density which is proportional to the 
power spectrum centered around $\omega_0$ is given by [$\hbar=1$] 
\begin{eqnarray}
u(\Omega,t) & = & \frac{\Omega}{4\pi} \int_{-\infty}^{+\infty} d\tau \;
\langle {\tilde{\xi}}^\dagger (t) {\tilde{\xi}} (t+\tau) \rangle 
e^{i(\Omega-\omega_0)\tau} \nonumber\\
& = & \frac{1}{2} \Omega {\bar{n}}(\Omega) + e^{-2\gamma(t-t_0)} \left [
u(\Omega,t_0) - \frac{1}{2} \Omega {\bar{n}} (\Omega) \right ] \; \; .
\end{eqnarray}

It is important to note that $t$ is the slow time variable which is well 
separated from the time-scale of thermal noise. The fluctuations of the noise
operator $\xi(t)$ is now explicitly determined by the nonequilibrium state
of the bath modes $\{c_\mu\}$ through its energy density $ u(\Omega,t)$
at each instant of time. In other words the instantaneous nonequilibrium
energy density distribution of fluctuating modes is related to the friction 
coefficient of these modes on the system degree of freedom through a dynamic
equilibrium. The classical version of the above equation can be recovered
in the high temperature limit 
($ {\bar{n}} (\Omega) =1/[\exp(\Omega/KT) -1] \simeq \frac{KT}{\Omega} $) to
obtain
\begin{equation}
u(\Omega,t) = \frac{1}{2}KT + e^{-2\gamma (t-t_0)} \left[ u(\Omega,t_0)
- \frac{1}{2}KT \right] \; \; .
\end{equation}

\noindent
This classical version was discussed earlier \cite{jcp} in the context of classical
kinetics of complex systems. Our quantum generalization is more relevant
to quantum optical situations as discussed in the next section.

\vspace{0.5cm}


\section{
Modified Bloch equations and transient resonance fluorescence}

\vspace{0.5cm}

We have discussed above a simple solvable model for a nonstationary quantum
Markov process and an appropriate generalization of the 
fluctuation-dissipation 
relation pertaining to this process. Two immediate consequences are evident.
The first one concerns the modification of decay of the Bloch vector components
in presence of relaxation of the intermediate bath modes. We show here that 
the decay is nonexponential in nature so long as the nonstationarity persists
following the sudden excitation. The second one centers around the explicit
time-dependence of diffusion coefficient due to nonstationarity implied in the 
fluctuation-dissipation relations (26). The transient noise spectrum of the
two-level system is therefore expected to bear this signature of time 
dependence. With this end in view we calculate the physical spectrum of the
two-level system  in contact with the nonequilibrium bath
driven by a near-resonant classical
monochromatic light field. The Hamiltonian of the coupled atom-field system
reads as follows;
\begin{equation}
H = H_0 + \hbar \left [ V \sigma_+ \; e^{-i\omega_c (t-t_0) }  \; + \;
V \sigma_- \; e^{i\omega_c (t-t_0) } \right ] \; ,
\end{equation}

\noindent
where $H_0$ is given by Eq. (1) and $V$ represents the amplitude of the 
classical pump field with frequency $\omega_c$. Proceeding as before, we 
obtain the Langevin equations for the slowly varying system operators as;
\begin{equation}
\left. \begin{array}{ccc}
{\dot{S}}_+ (t) & = & -(\Gamma-i\delta) S_+ + 2\xi^\dagger (t) S_z 
- 2iVS_z \\
{\dot{S}}_z (t) & = & -2\Gamma (S_z +\frac{1}{2} ) -iVS_+ + iVS_- + 
\xi (t) S_+  + \xi^\dagger (t) S_- \\
{\dot{S}}_- (t) & = & -(\Gamma + i\delta) S_- + 2\xi (t) S_z + 2iVS_z
\end{array} \right \} \; \; ,
\end{equation}

\noindent
where, $\delta ( =\omega_0 - \omega_c )$ is the detuning and
\begin{equation}
\left. \begin{array}{ccc}
S_+ (t) & = & \sigma_+ (t) e^{-i\omega_c (t-t_0)} \\
S_- (t) & = & \sigma_- (t) e^{i\omega_c (t-t_0)} \\
S_z (t) & = & \frac{1}{2} \sigma_z (t)
\end{array} \right \} \; \; ,
\end{equation}

\noindent
$\xi(t)$ is the noise operator as given by
\begin{equation}
\xi (t) = i\sum_\mu g_\mu \left [ c_\mu^s (t) + c_\mu (t_0) \; 
e^{(-i\Omega_\mu - \gamma) (t-t_0)} \right ] \; e^{i\omega_c (t-t_0)} \; \; .
\end{equation}

\noindent
The noise is characterized by 
\begin{eqnarray}
\langle \xi (t) \rangle_{NR} & = & 0 \nonumber \\
\langle \xi^\dagger (t) \xi (t') \rangle_{NR} & = & \Gamma \; {\bar{n}}(\nu)
\left [ 1+ r e^{-2\gamma(t-t_0)} \right ] \; \delta(t-t') \; \; .
\end{eqnarray}

\noindent
While considering the above equations we emphasize again the separation of
time scales $\gamma \ll \Gamma$.

As a next step, we construct the following Bloch equations for 
one-time averages 
from the Langevin equations (31) [ this requires the calculation
of averages like $\langle \xi^{\dagger} (t) S_{z} \rangle$  which include
nonstationary contribution involving
Eq.(34) as shown in Appendix-A ],
\begin{eqnarray}
\langle {\dot{S}}_+ (t) \rangle_{NR} & = & -(\Gamma -i\delta) 
\langle S_+ (t) \rangle_{NR} + 2\Gamma \; {\bar{n}} \; r e^{-2\gamma t} 
\langle S_+ (t) \rangle_{NR} - 2iV \langle S_z (t) \rangle_{NR} \; , \\
\langle {\dot{S}}_- (t) \rangle_{NR} & = & -(\Gamma + i\delta) 
\langle S_- (t) \rangle_{NR} + 2\Gamma ({\bar{n}}+1) r e^{-2\gamma t} 
\langle S_- (t) \rangle_{NR} + 2iV \langle S_z (t) \rangle_{NR} \; ,   \\
\langle {\dot{S}}_z (t) \rangle_{NR} & = & -2\Gamma \left [ \left \{ 
1- \left (2 {\bar{n}} + 1 \right ) r e^{-2\gamma t}  \right \}  
\langle S_z (t) \rangle_{NR} +\frac{1}{2} \right ]  \nonumber \\
& & -iV \langle S_+ (t) \rangle_{NR}  + iV \langle S_- (t) \rangle_{NR} \; .
\end{eqnarray}

The following comments should be made about the Eqs.(35-37) supplemented by 
Eq.(34) : 

\noindent
(i) The exponential term in Eq.(34) results in an  effective 
transient  modification of decay rates of all the Bloch vector components. 
In the long time limit  one, however, recovers the usual decay rates and the 
standard Bloch equations. It is also interesting to note that the two
polarization components $\langle S_+ \rangle$ and $\langle S_- \rangle$
decay at different rates in contrast to the usual case of equilibrium bath.

\noindent
(ii) The nonstationary contributions in the Bloch equations 
immediately assert that in absence of the driving fields ($V=0$) the decay 
of the polarization components is {\it non-exponential} in nature. This is 
reminiscent of what has been observed in the relaxation kinetics of classical 
complex systems where the influence of an initial nonequilibrium excitation
of others degrees of freedom of a complex system on the relaxation of a 
specific quantity of interest has been explored. 

\noindent
(iii) Although the noise correlation in Eq.(34) involves an exponentially decaying 
term, $\delta(t-t')$ makes the noise instantaneously correlated. This implies
that we consider here a broad-band reservoir instead of a colored bath. It is 
important to note that Lewenstein et.al. \cite{lewen} in a different context 
have considered earlier the atomic decay in presence of a colored reservoir. 
They have used the modified Bloch equations in
non-Markovian form (which involves exponentially decaying terms due to the 
finite response time of the reservoir) and shown how the effects of the
colored reservoir can be inhibited at large driving fields. Thus the 
origin of the exponential term in Eq.(34) is different. 

We now turn to the second issue, i.e., the calculation of the transient
resonance fluorescent spectra. Using matrix notation, the above three 
equations (35-37) for single time expectation values can be put in a compact 
form as;
\begin{equation}
\frac{d {\bf{u}} (t)}{dt} = {\bf M (t)} \; {\bf u} (t) + {\bf f}
\end{equation}

\noindent
where ${\bf u}(t)$ and ${\bf f}$ are the column vectors and are given by
\begin{equation}
{\bf u}  =  \left ( \begin{array}{c}
\langle S_+ (t) \rangle_{NR} \\
\langle S_z (t) \rangle_{NR} \\
\langle S_- (t) \rangle_{NR} \end{array} \right ) \; \; \; \; \; \;
{\rm and} \; \; \; \; \; \;
{\bf f} =  \left ( \begin{array}{c}
0 \\
-\Gamma \\
0 \end{array} \right ) 
\end{equation}
\noindent
with
\begin{equation}
{\bf M(t)}  =  \left ( \begin{array}{ccc}
-(\Gamma -i\delta)+2\Gamma{\bar{n}}r e^{-2\gamma t} & -2iV & 0 \\
-iV &  -2\Gamma+2\Gamma r (2 {\bar{n}}+1)e^{-2\gamma t} & iV \\
0 & 2iV & -(\Gamma +i\delta)+2\Gamma({\bar{n}}+1) r e^{-2\gamma t} 
\end{array} \right ) \; \; .
\end{equation}

Since the calculation of spectra rests on the evaluation of two-time 
correlation functions of the atomic operators it is essential to examine the
validity of quantum regression hypothesis in the present context. To this end
we note the following points.

The essential statistical properties of the intermediate bath modes are
contained in Eq.(34). This equation suggests a differential behavior in time
dependence of the two terms. First, the exponential time dependence is due to
the initial preparation at $t_0$ and subsequent relaxation at any time
( $t-t_0$ ) of the intermediate bath modes. So the nonstationary nature is
implied in this term. On the other hand the $\delta(t-t')$ term essentially
signifies the correlation of intermediate bath fluctuations, $\xi(t)$, at
times $t$ and $t'$. The presence of $\delta(t-t')$ ensures the broad band
nature and hence the Markov property of the bath. These statistical 
considerations, therefore, reveal that the dynamics of the two-level atom
is acted upon by a {\it nonstationary but Markovian stochastic process} due to
the intermediate oscillators. By using the Langevin description of Heisenberg
equation of motion, Lax \cite{lax} has proved that Markov property implies
regression theorem as well the converse. The Markov property is defined by 
the requirement that a Langevin force at time $t$ is uncorrelated to any 
information at earlier time $t'$. The regression hypothesis is a consequence
of this requirement. Lax has specifically shown \cite{lax} that Onsager's
original statement for an equilibrium situation is valid even for a  
nonequilibrium situation provided the system is Markovian. The validity of
regression hypothesis therefore implies that two-time correlation evolves
in the same way as one-time expectation value.

The equation for evolution of two-time correlation functions is then given by
\begin{equation}
\frac{d}{d\tau} {\bf v}(t_2,\tau) = {\bf M}(t_2,\tau) \; {\bf v}(t_2,\tau) + 
{\bf F} (t_2) \; \; ,
\end{equation}
\noindent
where
\begin{equation}
{\bf v} (t_2,\tau) = \left ( \begin{array}{c}
\langle S_+ (t_2+\tau) \; S_-(t_2) \rangle_{NR} \\
\langle S_z (t_2+\tau) \; S_-(t_2) \rangle_{NR} \\
\langle S_- (t_2+\tau) \; S_-(t_2) \rangle_{NR} \end{array} \right ) \; , \;
{\bf F}(t_2) = \left ( \begin{array}{c} 
0 \\
-\Gamma \langle S_-(t_2) \rangle_{NR} \\
0 \\ \end{array} \right ) \; \; .
\end{equation}

\noindent
The relevant correlation function $v_1(t_2,\tau)$ required for calculation
of spectra is the first component of the vector ${\bf v}(t_2,\tau)$ and is
given by
\begin{equation}
v_1(t_2,\tau) = \langle S_+(t_2,\tau) S_-(t_2) \rangle_{NR} \; \; .
\end{equation}
\noindent
We assume that the atom is initially in its ground state. The Eq.(41) is
then solved to calculate the correlation function (43). The details are given
in the Appendix-B. 

At this point it must be emphasized that since we are dealing with a 
nonstationary situation the standard steady state definition of spectrum is
not adequate to describe the transient spectral features. We therefore resort
to a non-steady state spectrum or the so called `physical spectrum' of the
emission from the two-level atoms, where the attention is focused on a
dynamic evolution of the spectrum following an abrupt excitation of the
atom and the intermediate oscillator modes. The main reason for studying the
time-dependent spectrum is that the familiar power spectrum which results from
Weiner-Khintchine theorem is not applicable to nonstationary processes.
Eberly and Wodkiewicz \cite{eberly} have shown that the suitably normalized
counting rate of a photo detector can be used to define a time-dependent 
spectrum. This definition allows the influence of the spectrum analyzer (
basically a Fabry-Perot interferometer, for example ) to be exhibited in the
spectrum so that the band limit of the measuring device is appropriately
incorporated which makes the spectrum free from ambiguities. It has also 
been emphasized [10] that when the instrumental width, $W$, is narrow enough
such that $W\ll\Gamma$ the spectrum appears to be qualitatively close to
Weiner-Khintchine spectrum. This transient spectrum has been used 
previously on several occasions \cite{bdeb}. Following Eberly and Wodkiewicz 
\cite{eberly} we define the time-dependent spectrum in terms of the time
correlation function $v_{1}(t_2,\tau)$ as follows;
\begin{equation}
S(t,\omega,W) = 2W \; Re \int_0^t dt_2 \; e^{-W(t-t_2)} \; \int_0^{t-t_2}
d\tau \; e^{(\frac{W}{2} - i\Delta)\tau} v_1 (t_2,\tau) \; \; .
\end{equation}

\noindent
Here $t$ is the elapsed time after the system and the intermediate
oscillator modes have been 
subjected to the initial excitation at $t = t_0 (= 0 )$, $W$ is the full width
of the transmission peak of the interferometer and $\Delta(=\omega - \omega_c)$
is the detuning, or frequency offset of the Fabry-Perot line center above
the frequency of the field $\omega_c$. It is important to note that the
time-dependent spectrum is expressed in terms of the two integrals. The first
integral is over the correlation time $\tau$ and is actually the counterpart of
Weiner-Khintchine spectrum band limited by the width $W$ of the measuring 
device, while the second one over $t_2$ takes into account of the 
nonstationarity which makes the spectrum $t$-dependent. 

Making use of Eq.(43) in Eq.(44), performing the integration over 
$\tau$ and $t_2$ and extracting the real part, we obtain numerically the 
time-dependent spectrum as discussed below. 

Since the excitation at $t=0$, prepares an initial nonequilibrium energy
density of the intermediate oscillator modes
which differs from its equilibrium value, the initial diffusion
coefficient $D(0)$ deviates from its stationary long time value $D(\infty)$.
This deviation is measured in terms of $r \left ( = \frac{D(0)}{D(\infty)} -
1 \right )$ [see Eq.(27)] or equivalently in terms of the ratio of the photon
numbers $\frac{ {\overline{n}} (\omega_0,t_0)}{ {\overline{n}} (\omega_0) }$.
Another quantity of interest is the rate of relaxation $\gamma$
of the nonequilibrium intermediate oscillator modes due to their coupling
to the thermal modes. Both $r$ and $\gamma$ contribute significantly to the
nonequilibrium version of fluctuation-dissipation relation [ Eq.(26)] which
is essential for understanding the influence of a nonequilibrium bath on
the transient fluorescence spectrum. In Fig.(1) we plot the physical spectra 
at three different times after the initial excitation at $t=0$ for the 
parameter set $r=0.4$, $\bar{n}=0.1$, $\Gamma=1.0$, $\gamma=0.1$ under 
resonance condition $\delta=0$ and instrumental
linewidth $W=4.0$ for a low value of field strength $V=2.5$.
The choice of parameter space is guided by the early work of
Eberly and Wodkiewicz \cite{eberly} on the physical spectra of resonance 
fluorescence. For a higher value of $V (=10.0)$ the expected three-peak 
Mollow structure appears as
shown in Fig.(2) ( all other parameters kept fixed ). The steady state is clearly
achieved around $t=10.0$. It is interesting to note that  in the spectra
showed in Figs.(1) and (2), the asymptotic condition corresponds to the
usual stationary bath case, presenting the narrowest line. The transient 
nature of broadening of the bands is due to the creation of a nonequilibrium 
bath mode energy density through $r$. Since $r$ is a measure of the excess
energy gained by the intermediate modes into which the system has to release
its energy, the system also draws some energy from these modes by virtue of 
the fluctuation-dissipation relation. The latter relation illustrates a
dynamical balance of inward flow of energy due to fluctuations from the
reservoir into the system and the outward flow of energy from the system
to the reservoir due to dissipation of the system mode. This nonstationary 
diffusion of fluctuations from the intermediate bath modes into the system 
leads to transient spectral broadening. This
persists so long as the nonstationarity remains. If the relaxation of these 
bath modes approaches the time scale of $\frac{1}{\Gamma}$ ( with increase of 
$\gamma$ ) the broadening effect no longer appears. In the next section we 
describe an experimental scheme to show how this can be realized within the
purview of a simple cavity QED experiment.

\vspace{0.5cm}


\section{Discussion of an experimental scheme and conclusions}

\vspace{0.5cm}

Based on a microscopic model for a nonequilibrium bath we have constructed the
modified Bloch equations which incorporate the effect of nonstationary
relaxation and calculated the transient resonance fluorescence
spectra of a two-level system driven by a near resonant strong classical
field. 

We now discuss a specific system presenting the transient broadening effect
studied in this paper. It is wellknown that the spectrum of the radiation 
emitted by a 
strongly driven system is considerably modified if the atoms are confined
in a cavity. For our purpose the dynamics may be conveniently described if
one considers a two-level Rydberg atom as a system contained in a cavity
( whose modes serve as the intermediate oscillator modes of the present model
). The cavity in turn is weakly coupled to the vacuum modes playing the role
of the equilibrium reservoir through the cavity losses. By sudden sweeping of
the resonance of the cavity it is possible to dump an appreciable amount of
energy on the cavity modes by changing the number of photons abruptly. ( The
tuning of cavity in studying the emission of strongly driven two-level
systems like Ba atoms into the modes of the cavity had been experimentally 
carried out both under adiabatic \cite{holm} and non-adiabatic \cite{zhu,gsa}
conditions in cavity QED experiments \cite{bdeb,holm,zhu}. ) 
This corresponds to the initial preparation of the
nonstationary state of the cavity modes by changing
$r\left [ =\frac{ {\overline{n}} (\omega_0,t_0)}{ {\overline{n}} 
(\omega_0) }, {\rm see}\; {\rm Eqs}.(26) \; {\rm and}\; (27) \right ]$
in such a way that the energy of these modes becomes suddenly higher
than the average energy. Once this nonstationarity is attained, the
effect of relaxation of the cavity modes on the emission of the strongly 
driven
( externally ) two-level atoms can be monitored by observing the transient
fluorescence spectrum. Since the atom-cavity interaction 
( $\Gamma$, say $\sim$ 100 MHz ) is strong
compared to the decay rate of the cavity modes 
( $\gamma$, say $\sim$ 20 MHz ), the separation of
time scales as required can be conveniently maintained. We also expect to
observe the nonexponential decay of emission of the excited two-level atom
into the modes of an optical cavity so long as the nonstationarity persists.

We now summarize the main conclusions of this study:

\noindent
(i) The microscopic model proposed here may serve as a simple solvable 
paradigm for a nonstationary quantum Markov process.

\noindent
(ii) We establish an appropriate generalization of the fluctuation-dissipation
relation and its classical correspondence pertaining to the above-mentioned
process.

\noindent
(iii) The origin of nonstationarity (or nonequilibrium nature of the bath)
lies in the creation and subsequent relaxation of an energy density
fluctuation distribution function of the intermediate bath modes following a
sudden excitation.

\noindent
(iv) Keeping in view of the systematic separation of time-scales involved in
the overall dynamics we have shown that the decay of the polarization and
population inversion components
of the Bloch vector is non-exponential so long as the nonstationarity
persists.

\noindent
(v) The nonstationarity of the bath results in  time dependence of the
diffusion coefficient which show up in the transient broadening of the
physical spectra of resonance fluorescence.

\noindent
(vi) We have outlined a simple experimental scheme within a cavity QED setup
to verify the aspects nonexponential decay and transient broadening of 
emissions from a strongly driven two-level system in a cavity.

Since the underlying model of relaxation employed here bears its origin
in complex coupled systems one may also
envisage guest-host systems embedded in a lattice ( where the immediate local
neighborhood of the guest comprises intermediate oscillator modes and the
lattice plays the role of a thermal bath ) as typical candidates for 
experimental realization of such transient fluorescent processes. We thus 
expect the model to be relevant in the context of single molecule
spectroscopy \cite{tamarat}.

\acknowledgments
Partial financial support by the Department of Science
and Technology, Govt. of India, is thankfully acknowledged. DSR is indebted 
to Professor G. S. Agarwal for discussions.

\newpage


\begin{appendix}

\section{Calculation of the average 
$\langle \xi^\dagger (t) S_z (t)\rangle_{NR}$ }

To calculate $\langle \xi^\dagger (t) S_z (t) \rangle_{NR} $ we proceed
as follows :

We have from Eq.(36)
\begin{eqnarray*}
\xi^\dagger (t) = -i\sum_\mu g_\mu \; \left [ \; c_\mu^{s\dagger} (t) +
c_\mu^\dagger (t_0) e^{(i\Omega_\mu -\gamma)(t-t_0)} \; \right ] \; 
e^{-i\omega_c (t-t_0)} \; \; .
\end{eqnarray*}

This can be written as
\begin{equation}
\xi^\dagger (t) = \xi^{s\dagger} (t) + \xi^{\dagger N} (t)
\; \; ,
\end{equation}

\noindent
where, $\xi^{s\dagger} (t) = -i\sum_\mu g_\mu c_\mu^{s\dagger} (t) 
e^{-i\omega_c(t-t_0)}$
represents the stationary ( long time ) fluctuation and
\begin{eqnarray*}
\xi^{\dagger N} (t) = -i\sum_\mu g_\mu c_\mu^\dagger (t_0) \; e^{ (i\Omega
_\mu - \gamma ) (t-t_0)} \; e^{-i\omega_c (t-t_0)}
\end{eqnarray*}

\noindent
denotes the fluctuations due to the coupling of the system 
with the relaxing modes. It is essential to note that because of the
relaxation $\xi^{\dagger N} (t)$ noise is nonstationary. It is
important to note that the separation of time scales of
$\xi^{s\dagger} (t)$ and
$\xi^{\dagger N} (t)$. $\xi^{s\dagger} (t)$ is much faster compared to
$\xi^{\dagger N} (t)$ and represents a Gaussian white noise. Also 
$\xi^{s\dagger} (t)$ and $\xi^{\dagger N} (t)$ are assumed uncorrelated 
\cite{jcp}.

\noindent
Thus we note
\begin{eqnarray*}
\langle \xi^\dagger (t) \; S_z (t) \rangle_{NR} = \langle 
\xi^{s\dagger} (t)\; S_z (t) \rangle_{NR} + \langle \xi^{\dagger N} (t)\;
S_z (t) \rangle_{NR} \; \; .
\end{eqnarray*}

\noindent
Since, $\xi^{s\dagger} (t)$ is much faster and describes a stationary process,
we write
\begin{eqnarray*}
\langle \xi^{s\dagger} (t)\; S_z (t) \rangle _{NR}=  \langle
\xi^{s\dagger} (t) \rangle_{NR} \;
\langle S_z (t) \rangle_{NR} = 0 \; \; .
\end{eqnarray*}

Thus we have
\begin{equation}
\langle \xi^\dagger (t) \; S_z (t) \rangle_{NR} =
\langle \xi^{\dagger N} (t)\; S_z(t) \rangle_{NR} \; \; .
\end{equation}

\noindent
Because of the exponential term $\exp[-\gamma(t-t_0)]$ in the expression
of $\xi^{\dagger N} (t)$, expression (A2)
describes a nonstationary average which cannot be equated to zero as shown
below.

\noindent
Making use of the identity
\begin{eqnarray*}
S_z (t) = S_z (t-\Delta t) + \int_{t-\Delta t}^t dt' \; {\dot{S}}_z (t')
\end{eqnarray*}

\noindent
and the expression for ${\dot{S}}_z (t)$, we get
\begin{eqnarray*}
S_z (t) & = & S_z (t-\Delta t) \nonumber \\
& + & \int_{t-\Delta t}^t dt' \; [ -2\Gamma 
\{ S_z (t') +\frac{1}{2} \} -iVS_+ (t') +iVS_-(t') +
\xi (t') S_+ (t') + \xi^\dagger (t') S_- (t') ] \nonumber \\
& = & S_z (t-\Delta t) \nonumber \\
& + & \int_{t-\Delta t}^t dt' \; [ -2\Gamma 
\{ S_z (t') +\frac{1}{2} \} -iVS_+ (t') +iVS_-(t') +
\xi^s (t') S_+ (t') + \xi^N (t') S_+ (t') \nonumber \\
& + & \xi^{s\dagger} (t') S_- (t') + \xi^{N\dagger} (t') S_- (t') ]
\; \; .
\end{eqnarray*}

\noindent
We then calculate the average
\begin{eqnarray*}
\langle \xi^{N\dagger} (t) S_z (t) \rangle_{NR} = \int_{t-\Delta t}^t
dt' \; \langle \xi^{N\dagger} (t) \xi^N (t') S_+ (t') \rangle_{NR}
\; \; ,
\end{eqnarray*}

\noindent
where we make use of the fact that $\xi^{s\dagger} (t)$ and 
$\xi^{\dagger N} (t)$ are uncorrelated \cite{jcp} and the operator
$S_z (t')$ at time $t'$ is not affected by fluctuation at a latter time
$t$. Following Bourret \cite{bourret,frisch}
and van Kampen \cite{van} we now make decoupling approximation ( which implies that
the correlation of fluctuations $\xi^N (t)$ is much short compared to the 
coarsed-grained time scale over which the average $\langle S_+ \rangle$
evolves in time ) to obtain
\begin{eqnarray*}
\langle \xi^{N\dagger} (t) \xi^N (t') S_+ (t') \rangle_{NR} =
\langle \xi^{N\dagger} (t) \xi^N (t') \rangle_{NR} \;
\langle S_+ (t') \rangle_{NR} \; \; .
\end{eqnarray*}

\noindent
We then use the fluctuation-dissipation relation for the 
$\xi^N (t)$ [ Eq.(37) ] to obtain
\begin{equation}
\langle \xi^{N\dagger} (t)\; S_z (t) \rangle_{NR} = \Gamma \; {\bar{n}} \;
r \; e^{-2\gamma t} \; \langle S_+ (t) \rangle_{NR} \; \; .
\end{equation}

We thus take note in passing that identification of a nonstationary part
$\xi^{\dagger N} (t)$  (which is not invariant under time translation) in
Eq.(A1) leads us to a non zero average like (A3). The other nonstationary
averages are similarly calculated.

\section{Calculation of two-time correlation functions}

The matrix ${\bf M}$ as defined in Eq.(40) may be rewritten as

\begin{equation}
{\bf M(t)} = {\bf M_0} + 2\Gamma \; r \; e^{-2\gamma t} \; {\bf M_1}
\end{equation}

\noindent
where
\begin{eqnarray}
{\bf M_0} & = & \left ( \begin{array}{ccc}
-(\Gamma -i\delta) & -2iV & 0 \\
-iV &  -2\Gamma & iV \\
0 & 2iV & -(\Gamma +i\delta)  
\end{array} \right ) \; , \\
\nonumber\\
\nonumber\\
{\bf M_1} & = & \left ( \begin{array}{ccc}
{\bar{n}} & 0 & 0 \\
0 &  2{\bar n}+1 & 0 \\
0 & 0 & {\bar{n}}+1 
\end{array} \right ) \; \; .
\end{eqnarray}

\noindent
The solution of Eq.(41) in terms of ${\bf M_0}$ and ${\bf M_1}$ is 
\begin{eqnarray}
{\bf v} (t_2,\tau) & = & \exp \left [{\bf M}_0 \tau - \frac{\Gamma}{\gamma} 
r \; e^{-2\gamma(t_2+\tau)} {\bf M}_1 \right ] \nonumber\\
& & \times \left \{ {\bf v}(t_2,0) + \int_0^\tau dt \; \exp\left [ -{\bf M}_0 t +
\frac{\Gamma}{\gamma} r e^{-2\gamma(t_2+\tau)} \; {\bf M}_1 \right ]\right \}
{\bf F}(t_2) \; \; .
\end{eqnarray}

\noindent
We assume that the atom is initially in its ground state. Then,
\begin{eqnarray*}
{\bf v}(t_2,0) = \left ( \begin{array}{c}
\frac{1}{2} ( 1+2\langle S_z \rangle_{NR} ) \\
-\frac{1}{2} \langle S_- \rangle_{NR} \\
0 \end{array} \right ) \; , \; {\bf u}(0) = \left ( \begin{array}{c}
0 \\
-\frac{1}{2} \\
0 \end{array} \right ) \; \; ,
\end{eqnarray*}
\noindent
where ${\bf v}(t_2,\tau)$ and ${\bf u}(t)$ are defined by Eqs.(42) and (39)
respectively.

\noindent
Defining a matrix ${\bf T}$ and a vector ${\bf g}$ as
\begin{eqnarray*}
{\bf T} = \left ( \begin{array}{ccc}
0 & 1 & 0 \\
0 & 0 & -\frac{1}{2} \\
0 & 0 & 0 \end{array} \right ) \; , \; {\bf g} = \left ( \begin{array}{c}
\frac{1}{2} \\
0 \\
0 \end{array} \right ) \; .
\end{eqnarray*}

\noindent
we can write
\begin{equation}
{\bf v} (t_2,0) = {\bf T} \; {\bf u}(t_2) + {\bf g} \; \; .
\end{equation}

\noindent
Hence from Eq.(B4), using Eq.(B5) we get
\begin{eqnarray}
{\bf v} (t_2,\tau) & = & \exp \left [{\bf M}_0 \tau - \frac{\Gamma}{\gamma} 
r \; e^{-2\gamma(t_2+\tau)} {\bf M}_1 \right ]  \nonumber\\
\nonumber\\
& & \times \left \{  \left[{\bf T} {\bf u}(t_2) + {\bf g} \right ] 
+ \int_0^\tau dt \; \exp\left [ -{\bf M}_0 t +
\frac{\Gamma}{\gamma} r e^{-2\gamma(t_2+\tau)} \; {\bf M}_1 \right ]\right \}
{\bf F}(t_2) \; \; .
\end{eqnarray}

\noindent
The solution of Eq.(38) is
\begin{eqnarray}
{\bf u} (t_2) & = & \exp \left [{\bf M}_0 t_2 - \frac{\Gamma}{\gamma} 
r \; e^{-2\gamma t_2} {\bf M}_1 \right ] \nonumber\\
\nonumber\\
& & \times \left \{ {\bf u}(0) + \int_0^{t_2} dt \; \exp\left [ -{\bf M}_0 t +
\frac{\Gamma}{\gamma} r e^{-2\gamma t} \; {\bf M}_1 \right ]\right \}
{\bf f} \; \; .
\end{eqnarray}

\noindent
We again define a matrix
\begin{eqnarray*}
{\bf K} = \left ( \begin{array}{ccc} 
0 & 0 & 0 \\
0 & 0 & -\Gamma \\
0 & 0 & 0 \end{array} \right )  \; \; ,
\end{eqnarray*}

\noindent
to write
\begin{equation}
{\bf F}(t_2) = {\bf K} \; {\bf u}(t_2) \; \; .
\end{equation}

\noindent
Using Eq.(B7) and Eq.(B8) we get from Eq.(B6) , the solution for two-time
correlation function in terms of the initial condition as follows;
\begin{eqnarray}
{\bf v} (t_2,\tau) & = & \exp \left ( {\bf M}_0 \tau - \frac{\Gamma}{\gamma} 
r e^{-2\gamma(t_2+\tau)} {\bf M}_1 \right )  \; \left[ {\bf T} \left [\left \{
\exp \left ( {\bf M}_0 t_2 - \frac{\Gamma}{\gamma} 
r e^{-2\gamma t_2} {\bf M}_1 \right ) \right \} \right. \right.
{\bf u}(0) \nonumber\\
\nonumber\\
& & \left. \left. + \exp \left ( {\bf M}_0 t_2 - \frac{\Gamma}{\gamma} 
r  e^{-2\gamma t_2} {\bf M}_1 \right )  \int_0^{t_2} dt \;
\exp \left ( - \left \{ {\bf M}_0 t - \frac{\Gamma}{\gamma} 
r  e^{-2\gamma t} {\bf M}_1 \right \} \right ) {\bf f} \right ]
+{\bf g} \right ] \nonumber\\
\nonumber\\
& &  + \exp \left ( {\bf M}_0 \tau - \frac{\Gamma}{\gamma} 
r  e^{-2\gamma(t_2+\tau)} {\bf M}_1 \right )  \left \{ \int_0^\tau dt 
\exp \left ( -\left \{ {\bf M}_0 t - \frac{\Gamma}{\gamma} 
r  e^{-2\gamma (t_2+\tau) } {\bf M}_1 \right \} \right ) \right \}
\nonumber\\
\nonumber\\
& & {\bf K} \left \{ \exp \left ({\bf M}_0 t_2 - \frac{\Gamma}{\gamma} 
r  e^{-2\gamma t_2 } {\bf M}_1 \right ) \; {\bf u}(0) \right. \nonumber\\
\nonumber\\
& & \left. +
\exp \left ( {\bf M}_0 t_2 - \frac{\Gamma}{\gamma} 
r  e^{-2\gamma t_2} {\bf M}_1 \right )  \; \int_0^{t_2} dt \;
\exp \left ( -\left \{ {\bf M}_0 \; t- \frac{\Gamma}{\gamma} 
r e^{-2\gamma t} {\bf M}_1 \right \} \right )  {\bf f} \right \} .
\end{eqnarray}

\end{appendix}

\newpage

\begin{center}
{\large{\bf Figure Captions}}
\end{center}

\vspace{0.5cm}

\begin{enumerate}
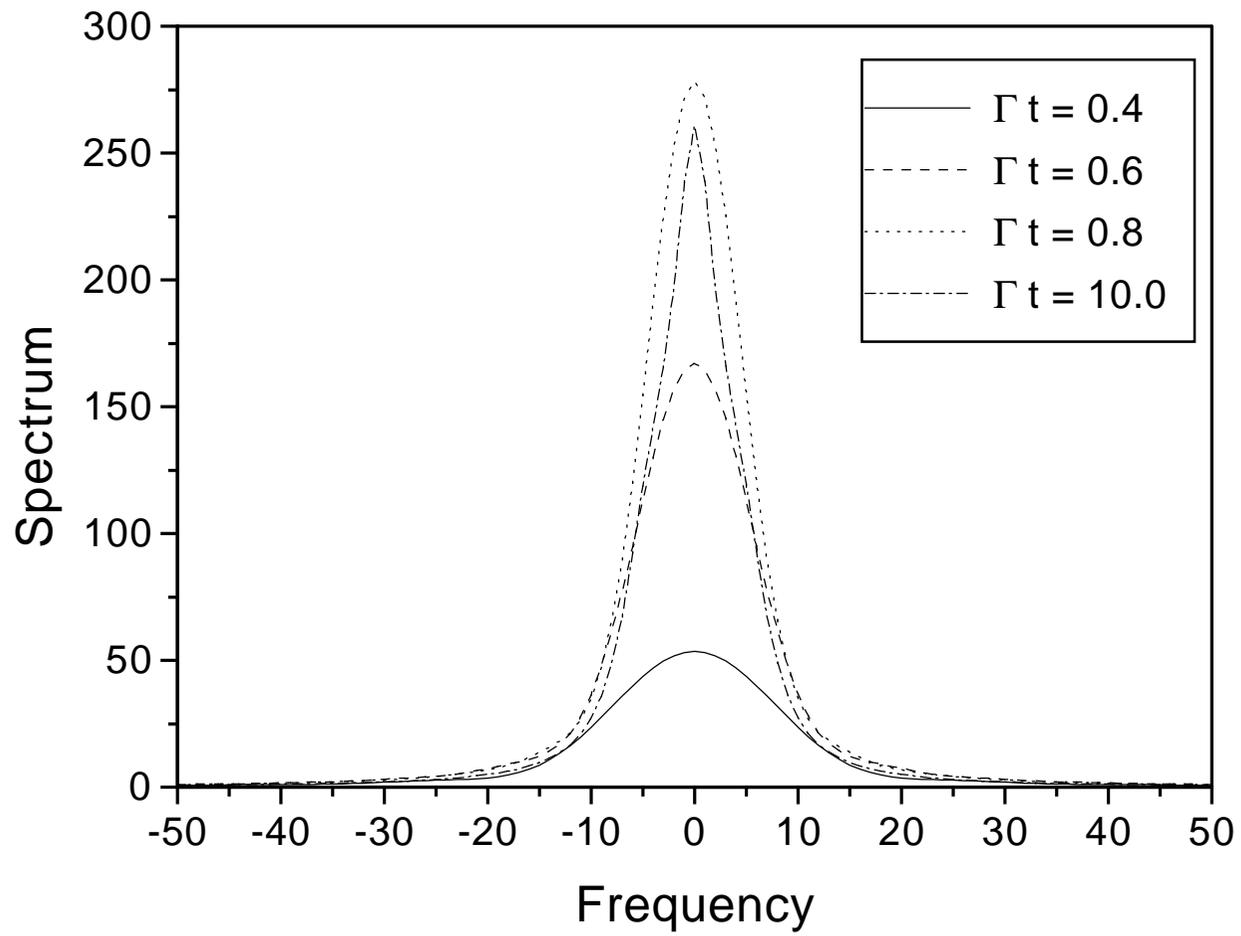
\item {\bf Fig.1} : Time-dependent resonance fluorescence spectra of the two-level 
system for different dimensionless times with W=4.0, $\delta=0.0$, ${\bar n}
=0.1$, r=0.4, $\gamma=0.1$ and V=2.5 (scales arbitrary).
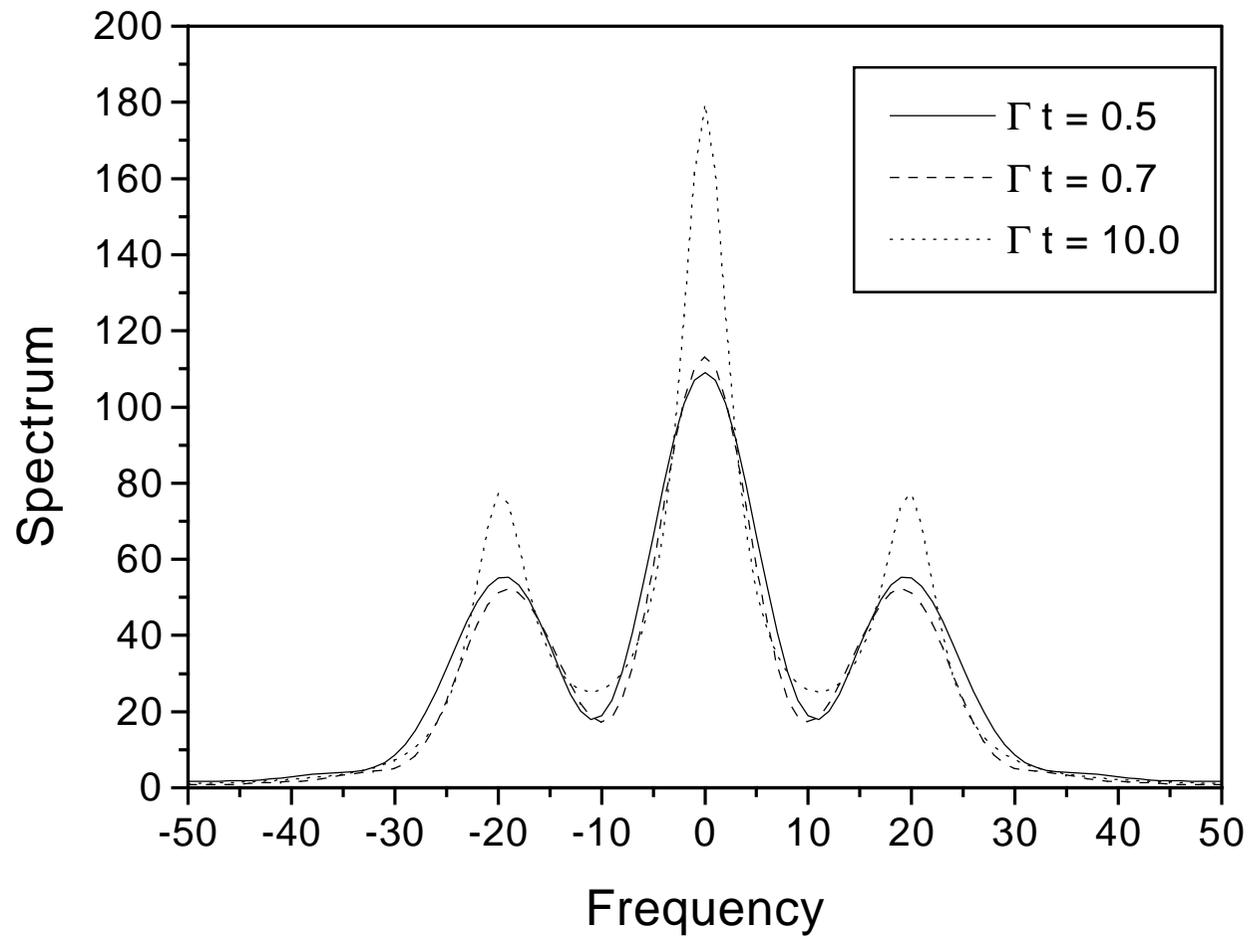
\item {\bf Fig.2} : Time-dependent resonance fluorescence spectra of the two-level 
system for different dimensionless times with W=4.0, $\delta=0.0$, ${\bar n}
=0.1$, r=0.4, $\gamma=0.1$ and V=10.0 (scales arbitrary).
\end{enumerate}
\end{document}